# Requirement Tracing using Term Extraction

Dr. Najla Al-Saati
Software Engineering Dept
College of Computer Sciences & Mathematics
Mosul, Iraq

Raghda Abdul-Jaleel
Software Engineering Dept
College of Computer Sciences & Mathematics
Mosul, Iraq

*Abstract*- **Requirements traceability is an essential step in ensuring the quality of software during the early stages of its development life cycle. Requirements tracing usually consists of document parsing, candidate link generation and evaluation and traceability analysis. This paper demonstrates the applicability of Statistical Term Extraction metrics to generate candidate links. It is applied and validated using two datasets and four types of filters two for each dataset, 0.2 and 0.25 for MODIS, 0 and 0.05 for CM1. This method generates requirements traceability matrices between textual requirements artifacts (such as high-level requirements traced to low-level requirements). The proposed method includes ten word frequency metrics divided into three main groups for calculating the frequency of terms. The results show that the proposed method gives better result when compared with the traditional TF-IDF method.**

## I. INTRODUCTION

The traceability of requirements was introduces mainly to manage and document the life of requirements. Its major objective is to maintain the activities of critical software development, for instance, the assessment of whether a software system has satisfied its definite set of requirements, the verification that all requirements have been employ by the end of the lifecycle, and the analysis of the impact imposed by the proposed changes on the system [1].

It is usually essential to follow the changes of requirements all the way through the lifecycle of software. All requirements should be validated in and at the end of each phase of the lifecycle. Traceability matrices are usually constructed to show the satisfaction of requirements by the design [2].

Generating traceability links (or traceability matrices) is fundamental to many software engineering activities [3]. But it is a time consuming, error prone, and mundane process. Most frequently, traceability matrices are built manually. When an analyst tries to trace a high level requirement document to a lower level requirement specification, he may have to look through M x N elements, where M and N are the number of high and low level requirements, respectively. Keeping in mind that there are very few tools available to assist the analysts in tracing unstructured textual artifacts, and those require enormous pre-processing [2].

Verification and Validation (V&V) and Independent Verification and Validation (IV&V) are used to ensure that the right processes have been used to build the right system. That is why it must be verified that the agreed processes and artifacts are directing the development in each phase of the life-cycle, in addition to ensuring that all requirements have been implemented at the end of the lifecycle. A requirements traceability matrix (RTM) is necessary for both of these [4][5].

The automatic generation of traceability links requires Information Retrieval (IR) techniques to reduce the time needed to generate the traceability mapping [3].

Requirements tracing usually enclose: document parsing, candidate link generation, candidate link evaluation, and traceability analysis. There are two commonly used measures for evaluating candidate link lists: *recall* and *precision*. In candidate link evaluation, the analyst investigates the candidate links and determines those that are actual (true links), and those that are not (false-positives, bad links). To achieve this, the analyst visually inspects the text of the requirements to find out the meanings of the requirements, compare them, and decide based on his believes which meanings are adequately close. This decision is based on human judgment and tolerates all the advantages and disadvantages that are related to it [4][5].

When tracing is finished, reports are generated by the analyst stating the high level requirements that do not have children and the low level elements that do not have parents (traceability analysis) [4][5].

## II. RELATED WORK

Many researchers have presented their work in requirement tracing during the last few years, such as:

In 2004 Hayes, et al. [5] designed RETRO to support the IV&V analyst in requirements tracing to find and evaluate candidate links.

Also in 2004, Sundaram, Hayes, and Dekhtyar [6] studied a mixture of IR methods used to solve the requirement traceability problem. They found that existing IR methods can be used in automating the generation of candidate links with minimal modification. And that the analyst's feedback information can considerably improve requirements tracing.

By 2006 Hayes, Dekhtyar, and Sundaram [4] inspected the efficiency of information retrieval methods in automating the tracing of textual requirements. They found that feedback from analyst improves final results via objective measures.

In 2007, Sundaram [2] assisted analysts in the traceability links generation process with information retrieval techniques for improving the quality of the generated links in addition to time saving.

Finally in 2010, Sundaram, et al. [3] stated that Information Retrieval techniques have been shown to aid in the automated generation of links through reduction of the time used in generating the mapping of traceability.

Researchers have successfully used techniques such as Latent Semantic Indexing (LSI), Vector Space Retrieval, and Probabilistic IR.

### III. REQUIREMENT TRACING

Requirements tracing is defined as the ability to describe and follow the life of a requirement, in both a forward and a backward direction, through the whole systems life cycle [2].

During the process of requirement gathering, the analyst has to clarify customer needs, conduct feasibility studies, specify a solution, and cross validates the specifications [7].

In large-scale projects, it is quite possible to miss or misinterpret some of the recognized requirements. More than 80% of the failures in large-scale mission-critical projects are caused by undetected problems in the early phases of the software development lifecycle [8]. An unobserved problem at the start of the project can continue all the way through to the deployed product; and becoming a latent defect or latent error [7].

Two sets of documents are typically created in the early phases of any software project:

- **Software Requirements Specification SRS**
  It is defined as "documentation of the essential requirements (i.e., functions, performance, design constraints, and attributes) of the software and its external interfaces. The software requirements are derived from the system specification [7]. SRS is a "binding contract among designers, programmers, customers, and testers," it includes different design views or paradigms for system design [9].

- **Software Design Description SDD**
  The design activity is used to identify the components of the software design and their interfaces from the Software Requirements Specification. The principal artifact of this activity is the Software Design Description (SDD) [9]. It is a "representation of software created to facilitate analysis, planning, implementation, and decision making". It is used as a medium for communicating software design information, and may be viewed as a blueprint or model of the system [7].

At the end of a requirements tracing process, a requirements traceability matrix (RTM) is generated [2]. RTM acts as a tool for indicating the way that the design and implementation elements deal with requirements throughout the whole software development lifecycle [7].

### IV. INFORMATION RETRIEVAL (IR) FOR REQUIREMENTS TRACING

Information retrieval (IR) is the process of discovery documents relevant to an information request in a collection of documents, usually a search query [7].

The main issue in IR is the determination of relevant documents in document collections given user-specified information needs. Most IR methods work by converting each document in the collection into a mathematical representation to capture the information content of the document, after that a comparison is conducted with similar representations of user information needs (queries). Nearly all IR methods are keyword-based: the document and query representations contain information regarding the importance of particular keywords found in the document [10]. There is a broad array of keyword-based retrieval models meant for document collections. The Boolean model is the simplest: a representation of a document is a Boolean vector identifying the keywords found in the document. A *Vector model* broadens the Boolean model by correlating each term in the document representation with a *weight* that signifies its understood importance to the document collection [11].

Documents and queries are represented as a vector of keyword weights. Formally, let $V = \{k_1,..., k_N\}$ be the vocabulary of a given document collection. Then, a vector model of a document $d$ is a vector $(w_1, ..., w_N)$ of keyword weights, where $w_i$ is computed as in Eq. (1) [10] [11].

$$w_i = tf_i(d) \cdot idf_i \quad \ldots\ldots\ldots\ldots\ldots\ldots\ldots\ldots\ldots(1)$$

Where
$tf_i(d)$ is the term frequency of the $i$th keyword in document $d$,
$idf_i$ is the *inverse document frequency* of the $i$th term in the document collection.

Term frequency is the number of term occurrences in the document and is usually normalized. The Inverse document frequency is computed using Eq. (2) [10][11].

$$idf_i = log_2\left(\frac{n}{df_i}\right) \quad \ldots\ldots\ldots\ldots\ldots\ldots\ldots\ldots(2)$$

*Where*
$df_i$ is the total number of documents containing the $i$th term in the document collection, and
$n$ is the size of the document collection.

The term significance is judged by how often this term is located in the document and by how discriminating the term is. That is, less frequent terms have more important presence for the document. A user query is also converted into a similar vector $q=(q_1,...,q_N)$ of term weights. In this model, given a document vector $d$ and a query vector $q$, the similarity between them is computed as the cosine of the angle between vectors $d$ and $q$ in the N-dimensional space as in Eq. (3) [10][11].

$$sim(d, q) = \cos(d, q) = \frac{\sum_{i=1}^{N} w_i \cdot q_i}{\sqrt{\sum_{i=1}^{N} w_i^2 \cdot \sum_{i=1}^{N} q_i^2}} \quad \ldots\ldots\ldots(3)$$

### V. EMPLOYD FILTERS

In this work, four filters are introduced to generate candidate link lists with relevance higher than one of the predefined levels: 0, 0.05, 0.2, and 0.25. This filtering acts as an assessment of the quality for the candidate link list. Having two candidate link list, say list X and list Y, with the same recall and precision, in that case if the true links show up at the top of list X compared with list Y, then obviously list X have preference to list Y from the analyst standpoint [2].

## VI. MEASURING THE EFFICIENCY

To evaluate the efficiency of IR techniques, recall (*R*) and precision (*P*) are used as the primary measures. recall measures if a method succeeded in finding all the high-low level requirement pairs that trace to each other, while recall indicates the number of additional pairs found by the method that do not trace to each other[6].

The computation of recall is done by dividing the total number of relevant retrieved documents by the total number of relevant documents in the complete collection, as in Eq.(6) [12].

$$R = \frac{\#of\_relevant\_retrieved}{\#\_relevant\_in\_collection} \quad \ldots\ldots\ldots\ldots\ldots\ldots\ldots\ldots(6)$$

The precision is calculated as the total number of relevant retrieved documents divided by the total number of retrieved documents, as shown by Eq.(7) [12].

$$p = \frac{\#of\_relevant\_retrieved}{\#\_retrieved} \quad \ldots\ldots\ldots\ldots\ldots\ldots\ldots\ldots(7)$$

## VII. TERM EXTRACTION

Term extraction forms an important issue in natural language processing; its goal is to extract sets of words with precise meaning in a collection of text. More than a few linguists considered these terms to be the base semantic unit of language. Automating term extraction comprises machine translation, automatic indexing, building lexical knowledge bases, and information retrieval [13].

Both supervised and unsupervised techniques have been used in earlier investigations to extract and distinguish terms. Nearly all researches aimed at locating the most significant set of terms from a domain corpus, to be precise, the set of superficial representations of domain concepts that better symbolize the domain for a human expert [14].

Term frequency in a corpus is a basic statistical property. This may then be compared to the frequency of the term in other corpora, such as balanced corpora or corpora from other domains. Basic frequency counts are integrated to compute co-occurrence measures for words. Co-occurrence measures are employed to estimate the propensity for words to appear together as multi-word units in documents, and to estimate the likelihood that units on either side of a bilingual corpus correspond under translation [15].

Term extraction can be used in this work to solve two issues:
- Finding high and low level requirements to create a common vocabulary. This is carried out using Statistical approaches, where all the terms are placed in a common vocabulary without any repetition.
- Using Statistical Term Extraction Metrics to calculate term weighting instead of TF-IDF in information retrieval.

## VIII. STATISTICAL TERM METRICS

In this work, ten standards metrics are proposed each as a measure instead of that used in the TF-IDF method, which was mentioned in Eq.(1). These metrics are divided into three main groups as explained in the following subsections [16].

Through the next subsections the following notations are used to symbolize equations: $tf_{ij}$ is the frequency of term *i* in each document *j*, *N* is the size of corpus. $w_i$ is the weight of term *i*.

### A. Term Frequency Based

The majority of term extraction algorithms base their results on some computation concerning term frequency.

*1) Corpus Term Frequency*

This metric is a solely term frequency metric, calculated over the entire corpus. It focuses on words that appear more often, except that it consequently favors large documents. Eq.(8) shows this calculation [16].

$$w_i = \sum_{j=1}^{N} tf_{ij} \quad \ldots\ldots\ldots\ldots\ldots\ldots\ldots\ldots\ldots (8)$$

*2) Logged Term Frequency*

Logarithms are considered as powerful modifiers of data, as they can reduce the range of values in a set. Logarithms are used to reduce the range of terms in any given document. This dampens the data, decreasing the distribution of frequencies as in Eq.(9) [16].

$$w_i = \sum_{j=1}^{N} \ln(tf_{ij} + 1) \quad \ldots\ldots\ldots\ldots\ldots\ldots\ldots (9)$$

*3) Document Term Frequency*

The maximum term frequency in a document is a unique metric, where the words that appeared most within their respective document are selected instead of summing together all the term frequencies. This is normalized, so as not to penalize words in short documents. This may provide new terms to the vocabulary by finding terms that appear often in one document, but not in any of the others. It favors unevenly distributed word frequencies, the calculation is done according to Eq.(10) [16].

$$w_i = \max_{1 \leq j \leq N} tf_{ij} \ldots\ldots\ldots\ldots\ldots\ldots\ldots\ldots(10)$$

### B. Normalization Based

Term normalization forms a standard metric for information retrieval; it is carried out by dividing the frequency of a term by the total number of terms in a document. When each document is normalized, the effect of size is removed, and each term frequency will form a percentage of another characteristic of the document, such as the document's term count [16].

*1) Document Terms Counts*

The widespread normalization of a document is carried out by dividing a term's frequency by the number of terms in a document [16]. Assuming Tj to be the total term count in document *j*, $w_i$ is calculated as in Eq.(11).

$$w_i = \sum_{j=1}^{N} tf_{ij} / T_j \quad \ldots\ldots\ldots\ldots\ldots\ldots\ldots (11)$$

*2) Document Maximum Frequency*

In this metric, the term frequency is divided by the most frequent term in a document, and the results are then summed up. The most frequent word gets a score of one for the document for which it is the most frequent term, in addition to any score it obtains by occurring in other documents. This has a similar effect to normalization because the score given to a term from any single document will not be greater than one, but the scores resulting from each document will be different than the scores after standard normalization. The weight *w* of term contributions is a ratio of the term frequency to the most common term $P_j$, rather than the frequency to the document size. Eq.(12) depict this [16].

$$w_i = \sum_{j=1}^{N} tf_{ij}/P_j \quad \text{...................................... (12)}$$

*a) Document Maximum Frequency & Term Average Frequency*

This metric also employs normalization according to the most frequent word in the document $P_j$, but here the average frequency that term *i* appears across as documents in the corpus is subtracted from Eq.(12). This is calculated as in Eq.(13) [16].

$$w_i = (\sum_{j=1}^{N} tf_{ij}/P_j) - \frac{\sum_{i=1}^{N} tf_{ij}}{N} \quad \text{................ (13)}$$

*3) Corpus Maximum Frequency*

The previous maximum frequency normalization technique can be further explored by using the most frequent term in the corpus. Being fixed, the corpus's most common term is a constant $P_c$. Results should be similar to the results of term frequency, if not exactly the same [16]. This metric is sometimes called corpus relativized. Eq.(14) shows this calculation.

$$w_i = \sum_{j=1}^{N} tf_{ij}/P_c \quad \text{.................................. (14)}$$

*a) Corpus Maximum Frequency & Term Average Frequency*

Referring back to the previous metric, the normalization was based on the most frequent term in the corpus, this metric is corpus relativized minus the average of TF as in Eq.(15) [16].

$$w_i = (\sum_{j=1}^{N} \frac{tf_{ij}}{P_c}) - \frac{\sum_{i=1}^{N} tf_{ij}}{N} \quad \text{.......................... (15)}$$

*C. Inverse Document Frequency*

The inverse document frequency measures desire words appearing in very few documents. It is used employed frequently in indexing; this is due to the fact that indexed documents in the corpus are in general varied, so a term that appears in few documents is a good identifier for those documents. Inverse Document Frequency together with term frequency assists in selecting words that occur repeatedly, but only in few documents [16].

*1) The TF-IDF*

Here, the term frequency is multiplied by the number of documents in the corpus, which is divided by the number of documents ($n_i$) that contain the term [16]. Eq. (16) shows the weight calculation.

$$w_i = \sum_{j=1}^{N} tf_{ij} * N/n_i \quad \text{.................................. (16)}$$

*2) Logged IDF*

This is similar to the TD-IDF measure, but here the term frequency is weighted more highly. The logarithm decreases the range of IDF values as in Eq.(17) [16].

$$w_i = \sum_{j=1}^{N} (tf_{ij}) * \ln\left(\frac{N}{n_i}\right) \quad \text{......................... (17)}$$

## IX. DATASETS

This work is validated using two NASA open source datasets. Both MODIS and CM-1 datasets are used here to assess the utilized techniques of IR. The MODIS dataset consists of 19 high level and 49 low-level requirements, where the CM-1 dataset contains 235 high-level requirements and 220 design elements. A manual tracing was done for both datasets for verification; these are referred to as "answer sets" or "theoretical true traces". There were 41 and 361 true links found for the MODIS and CM-1 datasets, respectively [6].

## X. EXPERIMENTAL RESULTS

Term Extraction is presented in this paper, along with a discussion of the Preprocessing techniques that are commonly used. First, the documents are parsed using the Statistical approach, stop words (words such as 'the' and 'of') are removed, and each remaining term is stemmed using Porter's algorithm [17], the term frequency is computed using ten word frequency metric rather than TF-IDF. In this paper the vector space model is used for Information Retrieval.

The four filters were used together with the metrics described previously using MODIS and CM1 datasets. The results are compared with those found by Sundaram et al. [6].

*A. First Dataset (MODIS) with Filters (0.2 and 0.25)*

In this section experiments are done using the MODIS Dataset and filters (0.2 and 0.25). Table (I) and (II) show the results of running the ten metrics for each filter. It was found that:

- **Filter 0.2**, <u>recall</u> value for all metrics improved, the best value was (68.2) achieved by the Document Term count metric and is labeled with (*) in Table (I). Best <u>Precision</u> is (23.7) in Term Frequency – Inverse Document Frequency metric.

- **Filter 0.25**, <u>Recall</u> improved for nearly all metrics except for Document Term Frequency, best value was (68.2) achieved by the Document Term count metric and is labeled with (*) in Table (II). Best <u>Precision</u> is (21.6) in Term Frequency – Inverse Document Frequency metric.

TABLE 1. RESULT OF METRICS IN MODIS DATASET WITH FILTER (0.2)

| Format | Term Weighting | Recall | Precision |
|---|---|---|---|
| XML[6] | TF_IDF | 19.5 | 21.6 |
| Statistical | Corpus Term Frequency | 65.8 | 13.5 |
| Statistical | Logged Term Frequency | 65.8 | 14.2 |
| Statistical | Document Term Frequency | 24.3 | 7.6 |
| Statistical | Document Terms Counts | 68.2* | 17.1 |
| Statistical | Document Maximum Frequency | 63.4 | 16.5 |
| Statistical | Document Maximum Frequency and Term Average Frequency | 65.8 | 17.0 |
| Statistical | Corpus Maximum Frequency | 65.8 | 13.7 |
| Statistical | Corpus Maximum Frequency and Term Average Frequency | 65.8 | 13.4 |
| Statistical | Term Frequency – Inverse Document Frequency | 34.1 | 23.7* |
| Statistical | Logged Inverse Document Frequency | 65.8 | 14.0 |

*\* best value*

TABLE II. RESULT OF METRICS IN MODIS DATASET WITH FILTER (0.25)

| Format | Term Weighting | Recall | Precision |
|---|---|---|---|
| XML[6] | TF_IDF | 19.5 | 32.0 |
| Statistical | Corpus Term Frequency | 65.8 | 16.0 |
| Statistical | Logged Term Frequency | 63.4 | 16.4 |
| Statistical | Document Term Frequency | 17.0 | 7.6 |
| Statistical | Document Terms Counts | 68.2* | 19.3 |
| Statistical | Document Maximum Frequency | 63.4 | 19.5 |
| Statistical | Document Maximum Frequency and Term Average Frequency | 63.4 | 19.6 |
| Statistical | Corpus Maximum Frequency | 65.8 | 16.1 |
| Statistical | Corpus Maximum Frequency and Term Average Frequency | 65.8 | 16.0 |
| Statistical | Term Frequency – Inverse Document Frequency | 19.5 | 21.6* |
| Statistical | Logged Inverse Document Frequency | 65.8 | 18.7 |

*\* best value*

TABLE III. RESULT OF METRICS IN CM1 DATASET WITH FILTER (0)

| Format | Term Weighting | Recall | Precision |
|---|---|---|---|
| XML[6] | TF_IDF | 97.8 | 1.5 |
| Statistical | Corpus Term Frequency | 97.7 | 1.0 |
| Statistical | Logged Term Frequency | 98.0 | 1.0 |
| Statistical | Document Term Frequency | 98.6* | 1.0 |
| Statistical | Document Terms Counts | 97.5 | 1.0 |
| Statistical | Document Maximum Frequency | 97.5 | 1.0 |
| Statistical | Document Maximum Frequency and Term Average Frequency | 97.5 | 1.0 |
| Statistical | Corpus Maximum Frequency | 97.7 | 1.0 |
| Statistical | Corpus Maximum Frequency and Term Average Frequency | 97.5 | 1.0 |
| Statistical | Term Frequency – Inverse Document Frequency | 98.6* | 1.0 |
| Statistical | Logged Inverse Document Frequency | 98.3 | 1.0 |

*\* best value*

TABLE IV. RESULT OF METRICS IN CM1 DATASET WITH FILTER (0.05)

| Format | Term Weighting | Recall | Precision |
|---|---|---|---|
| XML[6] | TF_IDF | 92.2 | 4.3 |
| Statistical | Corpus Term Frequency | 86.9 | 1.0 |
| Statistical | Logged Term Frequency | 87.5 | 1.0 |
| Statistical | Document Term Frequency | 93.0 | 1.1 |
| Statistical | Document Terms Counts | 86.1 | 1.0 |
| Statistical | Document Maximum Frequency | 86.9 | 1.0 |
| Statistical | Document Maximum Frequency and Term Average Frequency | 86.9 | 1.0 |
| Statistical | Corpus Maximum Frequency | 86.9 | 1.0 |
| Statistical | Corpus Maximum Frequency and Term Average Frequency | 86.1 | 1.0 |
| Statistical | Term Frequency – Inverse Document Frequency | 95.2* | 1.1 |
| Statistical | Logged Inverse Document Frequency | 92.7 | 1.1 |

*\* best value*

*B. Second Dataset (CM1) with Filters (0 and 0.05)*

Here, experiments are done using the CM1 Dataset and filters (0 and 0.05). Table (III) and (IV) show the results of running the ten metrics for each filter. It was found that:

- **Filter 0**, best Recall is (98.6) in Document Term Frequency and Term Frequency – Inverse Document Frequency metrics. Best Precision is (1.0) for all metrics as in Table (III).

- **Filter 0.05**, best Recall is (95.2) in Term Frequency – Inverse Document Frequency metric, Best Precision is (1.1) in Document Term Frequency, Term Frequency – Inverse Document Frequency and Logged Inverse Document Frequency metrics as in Table (IV).

In MODIS dataset, the **Recall** measure for both filters (0.2 and 0.25) showed better result for all metrics when compared to [6] except for Document Term Frequency in filter 0.25. Using the **Precision** measure, only Term Frequency – Inverse Document Frequency showed better results in filter(0.2), in filter 0.25 all of metrics showed less result than [6].

In CM1 dataset best value obtained in **Recall** measure was by using filter 0 and metrics (Logged Term Frequency, Document Term Frequency, Term Frequency – Inverse Document Frequency and Logged Inverse Document Frequency), which showed better results than [6], in filter 0.05 the Document Term Frequency, Term Frequency – Inverse Document Frequency, Logged Inverse Document

Frequency were better than [6]. In **Precision** all of metrics showed less result than [6].

In this work, focus was on improving recall at the cost of precision because high-recall, low-precision lists of links appear to be more preferable than high-precision, low recall links[4][5]. That is due to the fact that humans may be better at deciding if a specific pair of links in the list is a match than at finding new pairs of links in the document [5].

## XI. CONCLUSIONS AND FUTURE WORK

In this paper, the effectiveness of information retrieval methods in automating the tracing of textual requirements was examined. Ten metrics were evaluated and it was found that better recall can be achieved when compared to TF-IDF.

In this work, the vector space model was adapted for each of the metrics, in addition to the Statistical format. Porter Stemming Algorithm was applied using two open source datasets (MODIS and CM1).

Future work can carry on in several directions, such as the use of another technique in Information Retrieval (IR), as well as the vector space model. More methods can be sought to be employed other than term extraction to enhance results. Other datasets can also be used in this area.